\documentclass[twocolumn,prl,preprintnumbers,amsmath,amssymb]{revtex4}

\usepackage[english]{babel}
\usepackage{indentfirst}
\usepackage{graphicx}
\usepackage{amsmath}
\usepackage{amssymb}
\usepackage{amsthm}
\usepackage{mathrsfs}
\usepackage{lscape}
\usepackage{bm}
\usepackage{float}
\usepackage{longtable}
\usepackage{graphics}
\usepackage{color}

\newcommand{\be}
{\begin{eqnarray}}

\newcommand{\ee}
{\end{eqnarray}}

\usepackage{graphicx}
\usepackage{dcolumn}
\usepackage{bm}
\usepackage{times}

\usepackage{color}
\definecolor{rot}{rgb}{1,0,0}
\definecolor{blau}{rgb}{0,0,1}

\begin{document}

\bibliographystyle{unsrt}
\title{Observation of heavy-hole hyperfine interaction in quantum dots}
\author{P. Fallahi}
\author{S. T. Y\i lmaz}
\author{A. Imamo\u{g}lu}
\affiliation{Institute of Quantum Electronics, ETH Zurich, CH-8093
Z\"{u}rich, Switzerland}
\date{\today}

\begin{abstract}

We measure the strength and the sign of hyperfine interaction of a
heavy-hole with nuclear spins in single self-assembled quantum dots.
Our experiments utilize the locking of a quantum dot resonance to an
incident laser frequency to generate nuclear spin polarization. By
monitoring the resulting Overhauser shift of optical transitions
that are split either by electron or exciton Zeeman energy with
respect to the locked transition using resonance fluorescence, we
find that the ratio of the heavy-hole and electron hyperfine
interactions is $-0.09 \pm 0.02$ in two QDs. Since hyperfine
interactions constitute the principal decoherence source for spin
qubits, we expect our results to be important for efforts aimed at
using heavy-hole spins in solid-state quantum information
processing. The novel spectroscopic technique we develop also brings
new insights to the nuclear-spin mediated locking mechanism in
quantum dots.
\end{abstract}

\maketitle

Theoretical and experimental studies during the last decade have
established that hyperfine interaction with the quantum dot (QD)
nuclei constitute the principal decoherence mechanism for electron
spin qubits
\cite{Coish:2004,Petta:2005,Koppens:2006,Mikkelsen:2007}. An
interesting strategy to circumvent the leakage of quantum
information to the nuclear spin environment is to represent quantum
information with the pseudo-spin of a QD heavy-hole (HH): since HH
states are formed predominantly out of bonding p-orbitals of the
lattice atoms, it had been argued that the HH hyperfine interaction
should be negligible. Recently, it was shown theoretically that the
hyperfine interaction of a HH with the nuclear spins, while being
Ising-like, could in fact be comparable in strength to that of the
electron in strain-free GaAs QDs \cite{Fischer:2008}. However, the
majority of the experiments studying HH spins have been carried out
in highly strained InGaAs
QDs\cite{Brunner:2009,Gerardot:2008,Heiss:2008}; the strength and
nature of hyperfine interaction in these optically active QDs have,
to a large extent, remained unexplored \cite{Kurtze:2009}.

In this Letter, we present resonance fluorescence (RF) measurements
that directly reveal the relative strength of the HH-hyperfine
interaction in single-electron charged InGaAs QDs. To this end, we
generate a precise amount of nuclear spin polarization by dragging
the blue trion resonance using a non-perturbative laser field
\cite{Latta:2009}. We then measure the resulting Overhauser shift of
the QD transitions that are shifted either by the Zeeman energy of
the exciton (i.e. the red trion transition) or the electron (the
forbidden/diagonal transition) with respect to the blue trion
resonance. The nuclear spin polarization induced energy shifts in
these transitions are determined by the difference and the sum of
the Overhauser field seen by the electron and the HH, respectively.
Measuring the nuclear spin polarization induced shift of these two
transitions allows us to directly determine the ratio of the HH to
electron Overhauser shift to be $\eta=-0.09\pm0.02$ in two QDs and
$\eta=-0.10\pm0.05$ in a third QD.

\begin{figure}[h!]
\includegraphics[scale=1]{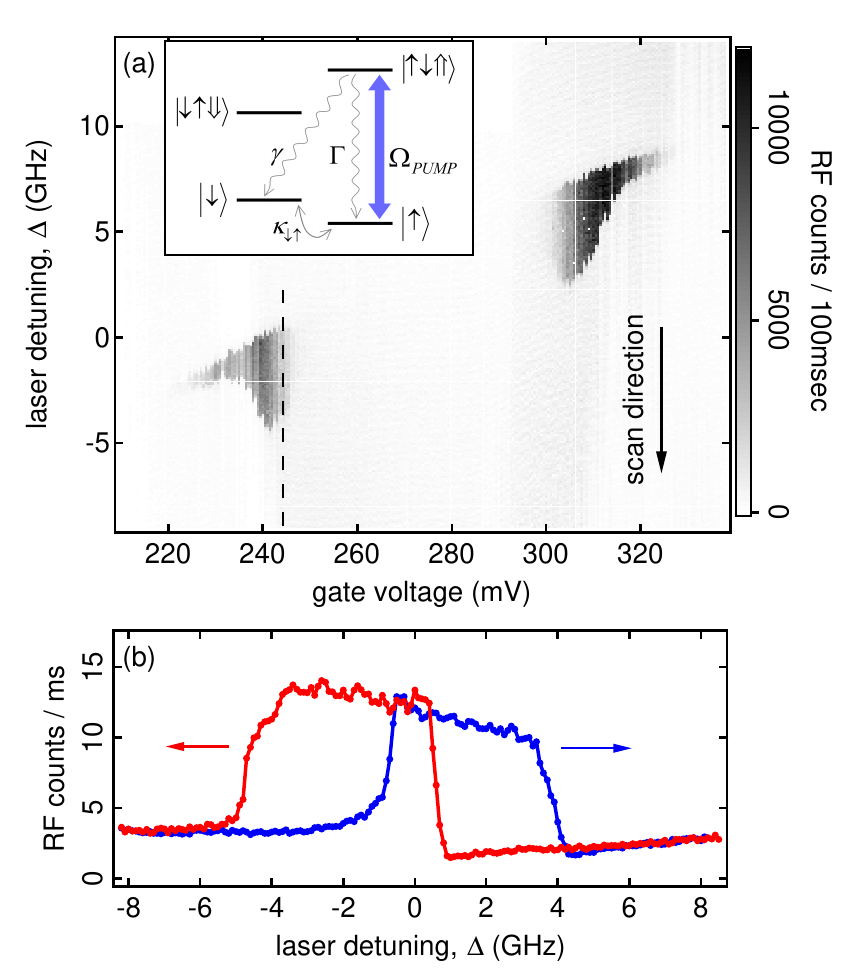}
\caption{\label{Fig1} (a) Resonance fluorescence  signal from the
blue trion transition as a function of gate voltage and pump laser
detuning, $\Delta$ at $B=4T$ and $P=P_{sat}/2$. Remainder of the
experiments are performed at the gate voltage indicated by the
dashed line, where the signal is reduced $\sim4$ times due to spin
pumping and a large line broadening due to dynamic nuclear spin
polarization is observed. Inset shows the energy level diagram for a
quantum dot charged with a single electron. (b) Cross section of (a)
across the dashed line opposite scan directions indicated by the
arrows. A total dragging range of $\sim8GHz$ is observed.
Interference with the laser background is partly responsible for the
change of resonance fluorescence (RF) counts along the dragging
range.}
\end{figure}

Our sample consists of charge-tunable InGaAs self assembled QDs
embedded in a Schottky-diode structure. All experiments are carried
out at 4.2 K with an external field in the Faraday geometry; the
applied gate voltage range is chosen to ensure that the QD is
single-electron charged. Figure~1 inset shows the relevant energy
levels in this regime where the QD optical transitions couple the
spin up $|\uparrow\rangle$ (spin down $|\downarrow\rangle$) ground
state to an optically excited blue (red) trion state
$|\uparrow\downarrow\Uparrow\rangle$
($|\uparrow\downarrow\Downarrow\rangle$). The two ground and excited
states are split by the Zeeman energies $|g_{e}| \mu _{B}B$ and
$|g_{h}| \mu _{B}B$. Due to heavy-light-hole mixing, the spin up
$|\uparrow\rangle$ (spin down $|\downarrow\rangle$) state also
couples to the other trion state
$|\uparrow\downarrow\Downarrow\rangle$
($|\uparrow\downarrow\Uparrow\rangle$), albeit with an oscillator
strength that is $\sim 10^{-3}$ times smaller; we refer to these as
the diagonal transitions.

Our experiments combine two recent advances in experimental QD spin
physics; namely the high-efficiency detection of resonance
fluorescence (RF)\cite{Yilmaz:2010,Vamivakas:2009,Flagg:2009} and
the possibility to lock a QD resonance to an incident laser
frequency via nuclear spin polarization \cite{Latta:2009}. Figure
\ref{Fig1}(a) shows the gate-voltage dependent RF signal from the
blue trion transition at $B=4T$ when the pump laser is scanned from
an initial blue-detuning ($\omega_p
> \omega_{t-b}^0$) to a final red detuning ($\omega_p < \omega_{t-b}^0$); here
$\omega_{t-b}^0$ is the trion resonance frequency in the absence of
nuclear spin polarization and $\omega_p$ is the pump laser
frequency. Reflected photons from the linearly polarized excitation
laser are blocked using a polarization suppression scheme
\cite{Yilmaz:2010}. The RF signal is strong at the edges of the
plateau where co-tunneling rate $\kappa_{\uparrow\downarrow}$ is
large, and disappears in the middle of the plateau due to spin
pumping \cite{Atature:2006,Dreiser:2008}. The characteristic
extension of the peaks in the direction of the laser scan is due to
dynamic nuclear spin polarization, i.e. the {\sl resonance dragging}
effect \cite{Latta:2009}. Experiments are performed at the gate
voltage indicated by the dashed line where the dragging range is
close to maximum. At this gate voltage the RF contrast is reduced
$\sim4$ times due to spin pumping. Figure\ref{Fig1}(b) shows laser
scans obtained at a gate voltage denoted by the dashed line in
Fig.~\ref{Fig1}(a) for two opposite scan directions: a total
dragging range of $\sim8$GHz is observed. We define the middle point
between the onset of forward and backward dragging to represent
$\omega_p - \omega_{t-b}^0 = \Delta = 0$. All other frequencies are
measured relative to this point. In the remainder of the
measurements we use the detuning $\Delta$ of the pump laser that is
slowly scanned across the blue trion transition, in either forward
or backward direction, as a knob for adjusting the amount of nuclear
spin polarization in the QD.

\begin{figure}[t]
\includegraphics[scale=1]{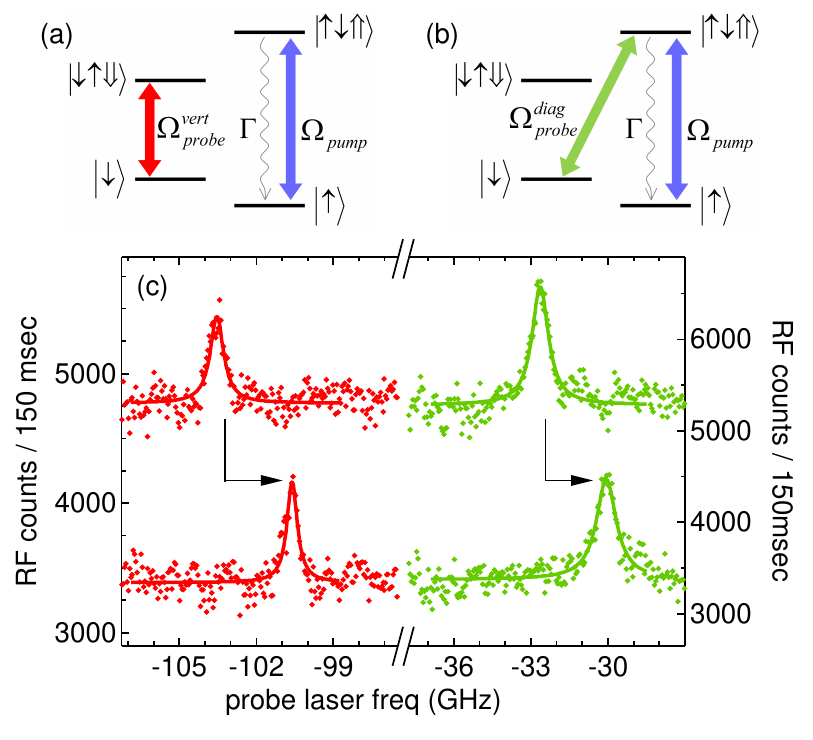}
\caption{\label{Fig2} (a)\&(b) Energy level diagrams showing the
pump and probe lasers. The probe laser re-pumps the spin into the
$|\uparrow\rangle$ state by driving the red trion (a) or the
diagonal (b) transition. (c) Resonance fluorescence (RF) signal
recorded as the probe laser is scanned across the diagonal (right)
or the red trion (left) transitions. Prior to the probe laser scan
the pump laser is scanned across the blue trion transition and
stopped at a detuning of $\Delta=0.5GHz$ (top) or $\Delta=-2.5GHz$
(bottom). Solid lines are Lorentzian fits. Peak positions are
shifted due to spin polarization induced by the pump laser.}
\end{figure}

Since the electrons in the relevant optically excited states form a
singlet, the nuclear-spin-polarization-induced change in the trion
Zeeman splitting is given exclusively by the Overhauser shift of a
single HH. To measure this HH Overhauser shift, we tune a strong
pump laser to create a precise amount of nuclear spin polarization
by dragging the blue trion transition from $\omega_{t-b}^0$ to
$\omega_{t-b}^0 + \Delta$. The pump laser which remains on resonance
throughout the dragging range, at the same time leads to a
substantial electron spin pumping into the $|\downarrow\rangle$
state, causing a reduction in the strength of the RF signal.
Subsequently, we scan a weak probe laser across the red trion and
the diagonal resonances: once the probe laser is on resonance with
either the red trion (Fig.~\ref{Fig2}(a)) or diagonal
(Fig~\ref{Fig2}(b)) \cite{Kroner:2008} transition, it pumps the
electron spin back to the $|\uparrow\rangle$ state, leading to a
partial recovery of the RF signal. The top traces in
Fig.~\ref{Fig2}(c) show the enhancement of the RF signal when the
probe laser is on resonance with the red trion (left) or the
diagonal (right) transitions, when the pump laser frequency was
fixed at $\Delta = 0.5 GHz$. When we scan the pump laser to a final
detuning of $\Delta =-2.5 GHz$, the resulting nuclear spin
polarization modifies both the red trion and the diagonal transition
resonance frequencies; the change in the red trion (diagonal)
resonance is given by $(-\delta E_e(\Delta)+\delta
E_{HH}(\Delta))/2$ ($(-\delta E_e(\Delta)-\delta
E_{HH}(\Delta))/2$), where $\delta E_{HH}(\Delta)$ ($\delta
E_{e}(\Delta)$) denotes the $\Delta$-dependent Overhauser shift seen
by a single QD HH (electron). Therefore, by measuring the shift in
the corresponding resonances using the probe laser
(Fig.~\ref{Fig2}(c) bottom trace), we determine the ratio
$\eta(\Delta)=\delta E_{HH}(\Delta)/\delta E_e(\Delta)$ of the HH
and electron Overhauser shifts to be $-0.09$. Our experiments reveal
that the HH hyperfine interaction has the opposite sign to that of
the electron.

\begin{figure}[t]
\includegraphics[scale=1]{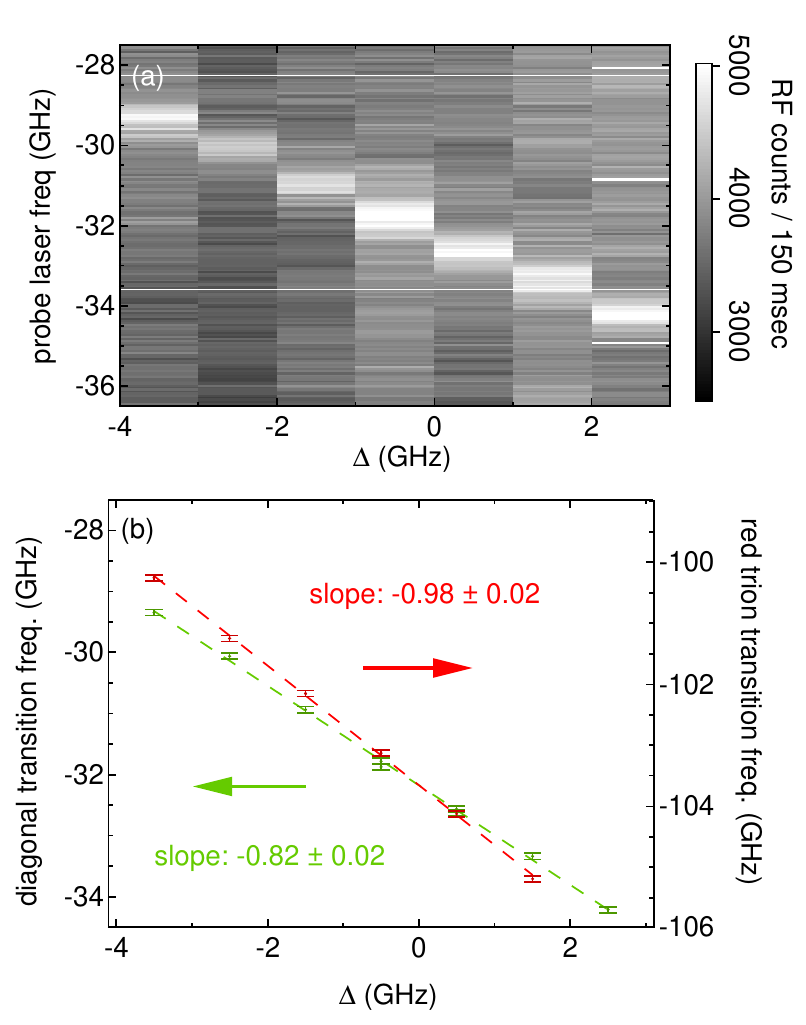}
\caption{\label{Fig3} (a) Resonance fluorescence (RF) signal as a
function of probe laser frequency scanned across the diagonal
transition and pump laser detuning, $\Delta$. The position of the
peak corresponding to recovered RF counts shifts due to nuclear spin
polarization induced by the pump laser. (b) (green points) The
position of the peaks in (a) extracted from Lorentzian fits to the
data. Red points correspond to peak positions extracted from similar
scans where the probe laser is scanned across the vertical
transition. Dashed lines are linear fits with slopes indicated on
the figure. Arrows point to the relevant axis.}
\end{figure}

The shift in the red trion and the diagonal transition frequencies
are extracted by fitting a Lorentzian lineshape to the resonantly
enhanced RF signal (Fig.~\ref{Fig2}(c)). To confirm that the ratio
$\eta(\Delta)$ is in fact independent of the amount of nuclear spin
polarization, we repeat the experiment for a range of different
$\Delta$ values. Figure~\ref{Fig3}(a) shows a series of probe scans
across the diagonal transition for different pump laser detunings.
The extracted transition energies are plotted in
Figure~\ref{Fig3}(b) (green dots). The red dots in
Fig.~\ref{Fig3}(b) show the shift of the red trion transition
measured using the same technique. Dashed lines are linear fits to
the extracted nuclear-spin-polarization-modified resonance
frequencies.

The remarkable linear fit to the data in Fig.~\ref{Fig3}(b) not only
shows that $\eta(\Delta) = \eta$ as we anticipated, but also that
both $\delta E_{HH}$ and $\delta E_e$ scale linearly with the
pump-laser detuning. This linear dependence allows us to simply
determine the relative strength of the HH hyperfine interaction as
$\eta = -(\alpha - \beta)/(\alpha + \beta)$, where $\alpha$
($\beta$) is the slope of the red-trion (diagonal) transition
depicted in Fig.~\ref{Fig3}(b). Using the measured slope of
$-0.98\pm0.02$ for the red line and $-0.82\pm0.02$ for the green
line we calculate $\eta=-0.09\pm0.02$.

\begin{figure}[b]
\includegraphics[scale=1]{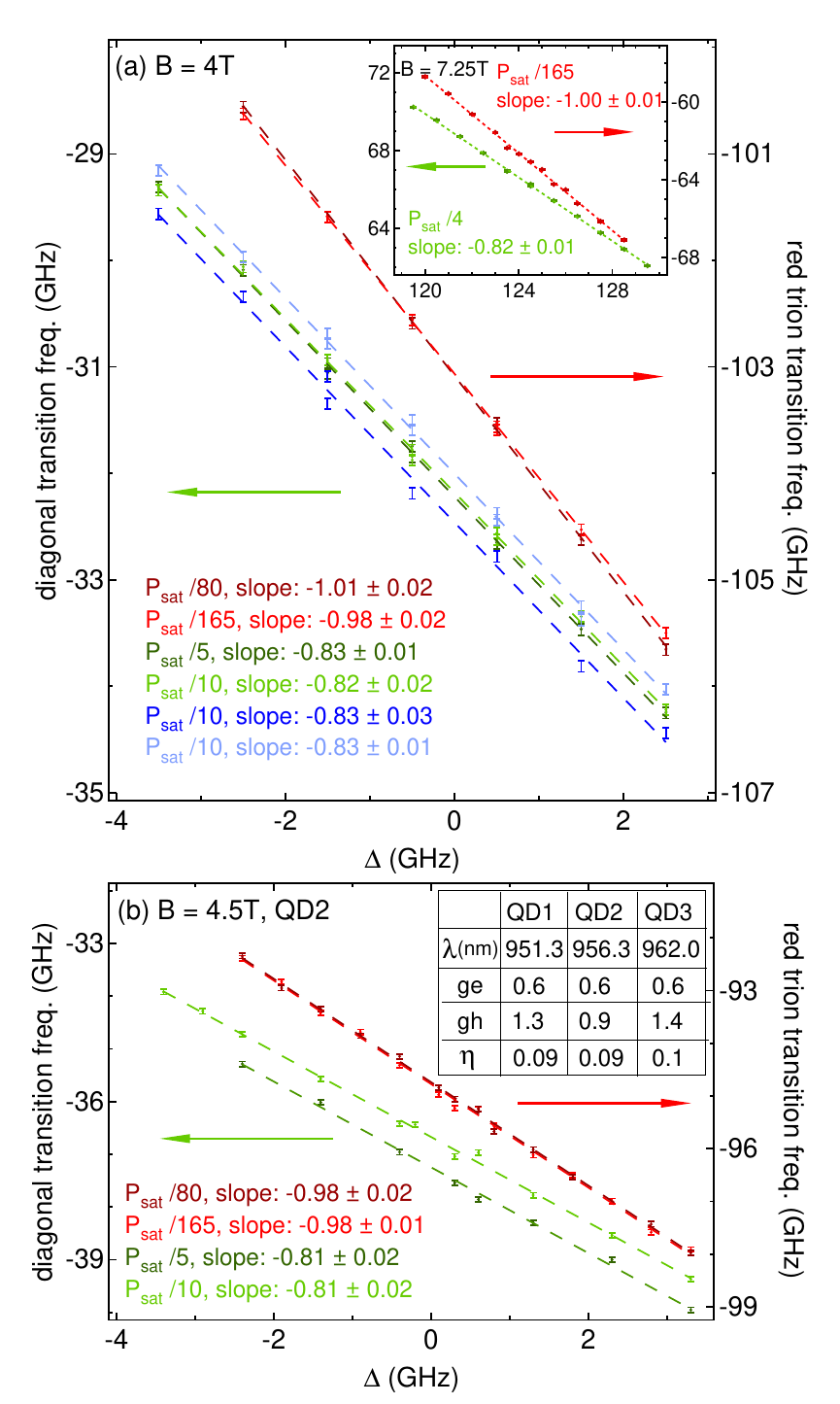}
\caption{\label{Fig4} (a) Transition energies of the red trion (red
points) and the diagonal transition (green and blue points) vs pump
laser detuning, $\Delta$, at $B=4T$. Probe laser power and slopes of
linear fits (dashed lines) are indicated on the figure. The dark and
light blue points are measured at gate voltages that are $\pm 1mV$
away from the one indicated on Fig.~\ref{Fig1}(a). The slopes do not
change with changing probe laser power or gate voltage. Inset shows
measurements at $B=7.25T$: the axes are the same as those on the
main figure. (b) Transition energies of the red trion (red points)
and the diagonal transition (green points) vs $\Delta$, measured on
a second quantum dot. The vertical shift between the two green lines
is due to a shift in the gate voltage, caused by fluctuations in the
quantum dot charge environment. The box compares the emission
wavelength, electron (ge) and heavy-hole (gh) g-factors and $\eta$
for 3 different quantum dots.}
\end{figure}

An accurate determination of $\eta$ requires that scanning the probe
laser across the red trion or the diagonal transition does not
modify the nuclear spin polarization that was created by the pump
laser. A clear signature of probe laser dragging, suggesting a
modification of the degree of nuclear spin polarization, is a change
in the pump laser induced RF signal, which constitutes the
background of the scans shown in figure~\ref{Fig2}(b) as the probe
laser is scanned across the transition. We choose the power of the
probe laser such that there is no measurable change in the
background. In addition, the linewidths of the recovered RF signal
match the QD transition linewidth at $B=0T$, in agreement with the
absence of probe induced dragging. Furthermore we repeated the
measurement at various probe laser powers; only when the laser power
is low enough does the slope become completely independent of laser
power. Figure~\ref{Fig4}(a) shows the Overhauser shifts of the
vertical and diagonal transitions, each for two different probe
laser powers indicated on the graph. The measured slopes at
different powers are indeed in excellent agreement. The oscillator
strength of the diagonal transition is much weaker than that of the
red trion transition, requiring higher probe laser powers to induce
resonant spin pumping. By repeating the diagonal transition
Overhauser shift measurements for two different gate voltages, $1
mV$ apart, around the gate voltage indicated by the dashed line on
Fig.~\ref{Fig1}(b) (blue traces on Fig.~\ref{Fig4}(a)), we have also
confirmed that the slopes we determined do not depend on the gate
voltage. The latter measurements also suggest that small changes in
the charging environment and the resulting changes in the confined
electron and HH wave-functions do not alter the ratio $\eta$. The
ratio $\eta$ also shows no appreciable dependence on the strength of
the external magnetic field (Fig.~\ref{Fig4}(a) inset).

Due to the large variation in HH g-factor and the positively-charged
trion confinement energy, it is generally believed that the confined
HH wave-function could change substantially from one QD to another.
To determine if these changes lead to a modification of HH-hyperfine
interaction, we have repeated our experiments on two other QDs.
Figure~\ref{Fig4}(b) shows measurements on a second QD which yields
$\eta=-0.09 \pm 0.02$. The ratio for a third QD, determined with a
factor 2.5 lower accuracy, yielded $\eta \sim -0.1$. Remarkably, we
find that the strength of the HH-hyperfine interaction in these 3
QDs to be almost identical, even though their HH longitudinal
g-factors vary substantially (Fig.~\ref{Fig4}(b)).

Despite the accurate measurement of the Overhauser shift of the HH
and the electron that we have demonstrated, it is not
straightforward to use our data to extract the actual HH hyperfine
interaction constant with high accuracy due to differences in the
confined electron and the HH envelope wave-functions. The exact
mechanism behind dragging of QD resonances is not well understood;
however, it is safe to assume that the underlying nuclear spin
polarization is mainly mediated by the electron \footnote[1]{We note
that W.Yang $\textit{et al.}$ \cite{Yang:2010} propose a dragging
mechanism based on the HH hyperfine interaction.}. The precise
magnitude of the HH Overhauser shift is therefore influenced by the
overlap between the electron and hole wave-functions and their
confinement length-scales. Repeating the experiments on different
QDs, particularly those with vastly different in-plane g-factors,
would yield further information about the sensitivity of $\eta$ to
the HH confinement.

The striking feature of our experiments is the (almost) perfect
linear dependence of the Overhauser shift on the pump-laser
detuning. What is even more remarkable is the fact that the slope of
the red trion energy shift is $\sim 1$, indicating that the
Overhauser shift of the blue trion transition satisfies ($-\delta
E_e(\Delta)+\delta E_{HH}(\Delta))/2 = \Delta + c$, where $c$ is a
constant much smaller than the bare optical transition linewidth. On
the other hand theoretical models proposed to explain dragging
showed a finite dependence of the absorption contrast on the bare
laser detuning \cite{Latta:2009,Yang:2010}, suggesting that the
amount of nuclear spin polarization has a non-trivial dependence on
the applied laser frequency. The experimental techniques developed
here could therefore help in identifying the physical mechanisms
underlying the dragging of QD resonances.

In summary, we have developed a new measurement technique combining
two recent advances in QD physics to determine the strength of the
HH hyperfine interaction. Our measurements on highly strained
self-assembled QDs indicate a coupling strength that is about a
factor of 2 smaller than what has been predicted theoretically for
strain-free GaAs QDs, and provide further support for efforts aimed
at using confined HH pseudo-spins as qubits in solid-state quantum
information processing.

We thank to Antonio Badolato for the sample growth and Martin Kroner
for many useful discussions. S.T. Y\i lmaz acknowledges financial
support from the European Union within the Marie-Curie Training
Research Network EMALI. This work is supported by NCCR Quantum
Photonics (NCCR QP), research instrument of the Swiss National
Science Foundation (SNSF). After completion of this work, we became
aware of related experiments by Chekhovic \textit{et al.}
\cite{Chekhovich:2010} on InP/GaInP QDs.

\end{document}